\documentclass[runningheads]{llncs}
\usepackage{graphicx}
\usepackage{xcolor}
\usepackage{url}
\usepackage{algorithmicx}
\usepackage{algorithm}
\usepackage[noend]{algpseudocode}
\usepackage[normalem]{ulem}
\usepackage{amsmath} 
\usepackage{booktabs}

\AtBeginDocument{%
  \providecommand\BibTeX{{%
    \normalfont B\kern-0.5em{\scshape i\kern-0.25em b}\kern-0.8em\TeX}}}
\newcommand{\MyNOTE}[1]{{\color{blue}\small$\langle\langle$ #1 $\rangle\rangle$}}    

\newcommand{\HIDE}[1]{}
\newcommand{\lmdh}[1]{{\fontfamily{qpl}\selectfont}}
\usepackage{comment}
\begin{document}
\title{Cooperative Multi-agent Approach for Automated Computer Game Testing}
%
%
\author{
Samira Shirzadeh-hajimahmood\inst{1}\orcidID{0000-0002-5148-3685} 
\and
I. S. W. B. Prasteya\inst{1}\orcidID{0000-0002-3421-4635} 
\and
Mehdi Dastani\inst{1}\orcidID{0000-0002-4641-4087}
\and
Frank Dignum\inst{2}\orcidID{0000-0002-5103-8127}}
\authorrunning{Shirzadeh-hajimahmood et al.}
%
\institute{
Utrecht University 
\email{\{S.shirzadehhajimahmood,s.w.b.prasetya,M.M.Dastani\}@uu.nl}
\and
Umeå University
\email{frank.dignum@umu.se}}
\maketitle              
\begin{abstract}
Automated testing of computer games is a challenging problem, especially when lengthy scenarios have to be tested. Automating such a scenario boils down to finding the right sequence of interactions given an abstract description of the scenario.
Recent works have shown that an agent-based approach works well for the purpose, e.g. due to agents' reactivity, hence enabling a test agent to immediately react to game events and changing state. 
Many games nowadays are multi-player. This opens up an interesting possibility to deploy multiple cooperative test agents to test such a game, for example to speed up the execution of multiple testing tasks. This paper offers a cooperative multi-agent testing approach and a study of its performance based on a case study on a 3D game called Lab Recruits.

\end{abstract}

\section{Introduction}



Modern computer games are often complex, with a huge, fine grained interaction space, and many interacting game objects 
that influence the way a game behaves. 
%
A common method for testing a game, to make sure that it behaves as the developers expect, is to use human players to play a computer game through various scenarios and report bugs \cite{ostrowski2013automated,iftikhar2015automated}. This technique is known as {\em play testing}. In addition to being costly, such a manual process is also unreliable, e.g. due to human fatigue. The tests also need to be repeated when some modifications are applied to the game.
The time and expense of having human testers repeat tests multiple times can be reduced by employing automated testing.

Agent-based approaches are among the recent methods studied towards achieving game testing automation \cite{ariyurek2019automated,albaghajati2020video,shirzadehhajimahmood2022online,prasetya2022agent}. Such an approach uses of a software agent that interacts with
the game under test by taking the role of a player. The agent verifies the system by observing its state after the interactions, checking them if they satisfy some specifications. In particular intelligent agents have properties such as reactivity and autonomy \cite{ch2007agent}.
Under the hood such an agent runs in continuous deliberation cycles, which makes it capable of responding immediately to changes in the environment (reactive), making them suitable to deal with the high interactivity of computer games.
Autonomy means that agents can make decisions based on their internal state and the information available to them, allowing the agents to perform actions independently in an environment over which they have control and observability.  



Many games are multi player. This opens up an interesting possibility of deploying multiple test agents to speed up testing, in particular when testing long scenarios that may take minutes or tens of minutes for each run. 
The overall duration of running a whole test suite can be quite significant, and therefore reducing would help in improving developers' productivity.
However, a multi-agent setup is more complex \cite{liu2022prospects}. A major challenge is cooperation. 
By cooperating, agents can improve the overall performance towards reaching common goals.
However, coordination may be needed, or else the agents will get in each other way and their performance will suffer instead.
Synchronizing the agent's information about their environmental perceptions can help, as each agent would then have the most recent information that other agents have, thus allowing it to make better decisions. Such synchronization does have its computation overhead though, and it is not always obvious to decide what information needs to be synchronized.

Compared to single-agent, the use of multi-agent for testing games is not a well explored area yet. One work we can mention is that of Schatten et al.
\cite{schatten2017automated} that presented a framework with which tests can be developed in a model-driven way. The framework implements
Belief-Desire-Intention (BDI) agents and allows organizational dynamics to be modelled. 
However, the work does not include a study on the subject. There are works such as \cite{gordillo2021improving,zheng2019wuji} that use multiple agents. The agents are trained with reinforcement learning to perform testing objectives, e.g. to interact and explore the game world as much as possible. These are individual agents that run in parallel to train a {\em common} model, or a population of common models. Arguably this is a form of multi-agent cooperation, but not in the sense that the agents directly cooperate with each other.

This paper aims to contribute an investigation into the subject of multi-agent game testing. A cooperative multi-agent testing approach will first be presented. At the moment, the approach is aimed at games with world exploration and puzzle elements. 
Next, the paper presents a study based on a 3D game called Lab Recruits to investigate whether, and 
    how much, the use of cooperative agents can actually speed up testing.

The paper is structured as follows. Section \ref{multi-agent-setup} discusses the general idea of using cooperating agents for testing a game and some of the key issues of such a setup.
Section \ref{problem setup} introduces the concept of testing task in game testing and some typical challenges of automating such tasks.
Then, Section \ref{sec:multi-agent-algorithm} presents our multi-agent testing algorithm.
Section \ref{sec:experiment} presents a set of experiments to study the performance the multi-agent setup.
Section \ref{sec:related-work} discusses some related work, and finally Section \ref{sec:concl} concludes and mentions some future work.

\section{Multi Agent Setup}\label{multi-agent-setup}

\begin{figure}[h]
\centering
\includegraphics[scale=0.4]{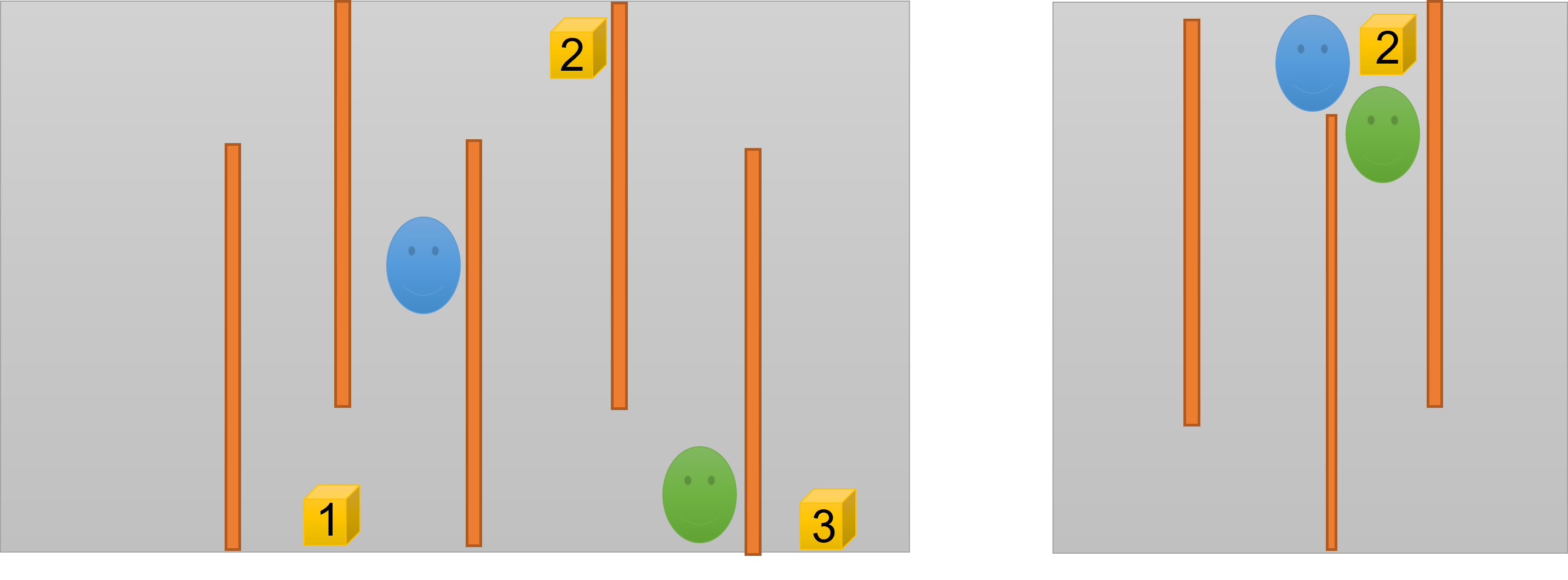}
\caption{\rm\em Two simple levels of a hypothetical game.}
\label{fig.SetupExample}
\vspace{-8mm}
\end{figure}

Consider a simple game level in Fig. \ref{fig.SetupExample}-left, containing three yellow treasure boxes ($T_k$). Picking up the boxes awards the player some points, which are necessary to complete the game. Imagine that the tester wants to verify that these boxes are indeed reachable from the game initial state and that their logic is correct. So, the tester has {\em multiple} 'testing tasks' that need to be done: one task for every $T_k$. For convenience, let's use $T_k$ to also denote the task.
%
In a single test-agent setup, a single agent will do all testing tasks, one at a time. For $T_k$ the test agent will first need to find corresponding box, and hence verifying its reachability, and then verify whether the points rise after picking it up is correct. 

In a multi agent setup, we have multiple test agents that work on a set $\mathcal{T}$ of testing tasks.
We will focus on the setup where the agents target the same instance of the game under test (GUT), exploiting its multi player capability\footnote{As opposed to performing parallel testing against multiple instances of the GUT.}.
The agents have to figure out how to divide the work. This is less trivial than it sounds.
The agents can choose to do different tasks in parallel.
For example, agent Blue in Fig. \ref{fig.SetupExample}-left can take $T_1$ and agent Green can take $T_3$. This division of tasks is ideal because $T_1$ is closest to agent Blue, and $T_3$ is closest to agent Green. However, if the area is large and the agents do not have full visibility on the world, it is not possible to know upfront which division of tasks is the best. 
%
Alternatively, we can have multiple agents working in parallel on the same task. For example, suppose  agent Blue decides to do $T_3$. Without full visibility, it does not know where $T_3$ is, and may end up searching the entire game world before it finally finds $T_3$. Having agent Green to also work on $T_3$ would speed up the task, as it happens to be close to it.


In our study we will focus on autonomous agents that dynamically chose the task they want to do, as opposed to having a central process that allocate the tasks. By 'dynamic' we mean that while an agent is exploring the game it on the fly decides on taking or ignoring a task. 


\subsubsection*{Cooperation.}
We will employ information synchronization and coordination as the form of cooperation in our setup.
Sharing and synchronising the agents' observation may save the effort that each agent would otherwise need for exploration and data collection. 
E.g. in the previous example where agent Blue targets $T_3$, if another agent shares the location of $T_3$, agent Blue might be able to find a path to it faster.
In the study in Section \ref{sec:multi-agent-algorithm} we will look at several options in the degree of information sharing.
For example, agents may only share partial knowledge about the game. 
Such characteristics would have an impact on how successful the testing task is executed. 


Having multiple agents running simultaneously may require some coordination.
For example, if two agents run into one another in a narrow corridor e.g. as in Fig. \ref{fig.SetupExample}-right they will get stuck.
Generally, executing a task may require resources critical to the task, which should not be shared, or else the task may be disrupted or even cannot be completed. The narrow corridor in Fig. \ref{fig.SetupExample}-right is such a resource.
In a game there are often objects that act as critical resources towards certain testing tasks, for example keys, switches, or even healing items. Access to such resources needs to be coordinated, e.g. through locking.

On top of the aforementioned challenges, the agents also have to deal with the game logic itself, which can be a challenge of its own. We will discuss more on this in the next section.
And later, in 
Section \ref{sec:multi-agent-algorithm}, we will discuss our multi-agent algorithm. It features the aforementioned dynamic task selection, information sharing, and locking of key resources.

\section{Problem Setup} \label{problem setup}
In this section, we describe a general game setup and outline the challenges of this setup. We abstractly treat a game under test as a structure  $G = (A,O)$
%
%
where $A$ is a set of players/agents and $O$ is a set of game objects that have physical locations in the game. 
We will use some or all agents to test the game, so in our setup they are test agents.
Game objects have their own state. Some objects are interactable. Some may be hazardous. Objects such as doors are called {\em blockers}; these can block the agent's access to an area.
Dealing with blockers is important for testing tasks that aim to verify the reachability of a state.

Interacting with an object $o$ may change the state of other objects. 
To provide challenges for players, game designers often make blockers' logic non-trivial. For example, it may require an interaction with another object, called an {\em enabler}, to unblock a blocker. The location of the enabler can be far, and not easy to find. Or it can be placed in an area that is guarded by another blocker, that in turn requires its own enabler to be found and activated, and so on.
Across different games there are different types of enablers. E.g. a toggling switch toggles the state of associated blockers (from blocking to unblocking, and the other way around). A one-off switch can only be used once. A key must be picked up, and brought to a blocker to be used, and so on. In this paper, we will restrict ourselves to toggling switches. 

Each agent $a {\in} A$ is assumed to only be able to {\em observe} objects and parts of the game/environment that it {\em physically can see}. 
Also, the agent does not know upfront how to how to solve complex tasks, such as unblocking a blocker. Examples of primitive actions available to an agent are {\em interacting} with an object (if the agent is close enough to it) 
and {\em moving} in any direction for a small distance. 
In our setup we will assume that 
a high level navigation action is also available to the agent, to auto-navigate to a specified location. This can be achieved e.g. by leveraging a game testing framework like iv4XR \cite{prasetya2020aplib} that comes with a path planning module.


The agent can take one or more testing tasks to do. 
Abstractly, such a task is be formulated as follows: 

\begin{equation} \label{eq.testingtask}
   Testing Task  = ( o,\psi,S)
    \end{equation}

Where $\psi$ is 
a predicate to hold on an object $o$ and S is a stop condition to terminate the test. 

We will focus on tasks whose objective is to verify whether the state of some game object $o$, characterized by the predicate $\psi$, is {\em reachable}. 'Reachable' means that there exists a sequence of agent's actions, starting from the game initial state, that leads to a state where $\psi$ is true. This is simple but still represents many useful testing tasks.
For example, if $o$ is a blocker, $\psi$ can specify that $o$ is open and visible to the agent. Verifying that $\psi$ is reachable (thus that $o$ can be unblocked, by activating some game logic) implies that the area that $o$ guards is thus reachable for the player, which is important if the area is critical towards the game's story line. 


We treat a testing task as a goal that a test agent wants to automatically achieve/solve (and thus providing test automation). 
To do this, the agent needs to {\em search} for a sequence of interactions that reach a state satisfying $\psi$ (thus confirming its reachability), while respecting the game rules. 
Finding such a sequence is usually not trivial. 
Different heuristics are defined for this, which we will explain later. The parameter $S$ in  a testing task is used to terminate the task, e.g. based on time budget, after which the goal $\psi$ is judged as unreachable.

\section{Multi Agent Testing Algorithm} \label{sec:multi-agent-algorithm}
 We propose a cooperative multi-agent approach to solve the given set of testing tasks automatically. The details are provided in this section. The approach has two main algorithms. Algorithm \ref{alg:multiAgents} takes a set $\mathcal{T}$ of testing tasks that the developers have specified and need to be carried out.
The algorithm introduces some variables to keep track of the status of the tasks, e.g. which ones are completed, then runs $N$ agents. It also deals with information sharing. Algorithm \ref{alg:solver} called \Call{Solver}{} defines how each agent chooses testing tasks and how it proceeds to complete a chosen task.

\begin{algorithm}
\small
\caption{\em It gets a set of tasks $\mathcal T$ and run $N$ test agents.}\label{alg:multiAgents}
\begin{algorithmic}[1]
\Procedure{CooperativeAgents}{$\mathcal{T}$} 
\State $toDo = \emptyset$ 
\State $done  = \emptyset$ 

\State $ agent_{1}.\Call{solver}{{\bf H}_1}  \; \parallel \; ... \; \parallel \; agent_{N}.\Call{solver}{{\bf H}_N}$ 
\State \ \ \ \ $\parallel \Call{Sync}$
\State \ \ \ \ $\parallel$ $  done{=}\mathcal{T} \vee budget \leq 0\ \rightarrow \ $ {\bf terminate} \Comment{terminate the whole procedure}
\EndProcedure
\end{algorithmic}
\end{algorithm}

The agents run in principle in parallel, each will repeatedly select a task from $\mathcal{T}$ and try to complete it. Certain coordination will be needed; we go into more detail about that later.
The notation $P \parallel Q$ in Algorithm \ref{alg:multiAgents} denotes a parallel execution of processes $P$ and $Q$. A process is a program that sequentially executes primitive actions. For an agent $a$, the process $a.\Call{solver}{}$ runs the algorithm \Call{solver}{} for the agent $a$. 
The process $\Call{Sync}$ is responsible for regular sharing and synchronization of agents' observations with each other.

\HIDE{
We will assume a simple synchronous model of parallel execution \MyNOTE{why?}. This means that primitive actions from $P$ and $Q$ can execute in parallel, but they need to synchronize after every action.
More precisely, $P \parallel Q$ means executing the following loop:
\[ 
\begin{array}{l}
{\bf while} \; true  \\
\ \ \ \ a \gets \mbox{the next enabled action of} \; P \; \mbox{, else take skip} \\
\ \ \ \ b \gets \mbox{the next enabled action of} \; Q \; \mbox{, else take skip} \\
\ \ \ \ a \parallel b
\end{array} 
\]
where $a \parallel b$ is the parallel execution of actions $a$ and $b$.
}

The agents attempt to complete as many tasks from $\mathcal{T}$ as possible according to their task selection policy, as long as the budget that was given is not used up. Variables $toDo$ and $done$ are empty at the beginning and will be updated during the test execution.  When an agent discovers an object $o$ which is targeted by a testing task $T \in \mathcal{T}/done$, this $T$ is added to the set $toDo$ to keep track of uncompleted tasks whose location of their target objects are known. When $T$ is completed, it is moved from $toDo$ to the set $done$. 
\HIDE{
 \MyNOTE{
The concept of 'satisfied' and 'solved' are not firmly defined. You mentioned something about it in S3. 

Is a testing task is satisfied when it finds a sequence of interactions that reach a state satisfying $\psi_o$ ?

The task is solved: is that synonymous to satisfaction?

What if the task times out: do we cal this: "the testing task fails"?

How about task 'completion'? When a task fails, do you consider that to be completed? 

And finally, does $done$ contains only solved tasks, or completed tasks?
}
} 
This is done in Algorithm \ref{alg:solver}
\Call{Solver}{}.

In \Call{Solver}{},
the agent that runs it continues attempting to solve tasks until there is no task left and there is no unexplored area left. To choose a task from the $toDo$ the agent uses an assigned selection heuristic $selectH$. E.g. the heuristic might favor high-valued tasks.
We will go for a scheme where each task is worked on by at most one agent. This is enforced by removing the selected task from $toDo$.

Suppose $T = (o,\psi,S)$ is the selected task. To check if $\psi$ already holds the agent needs to be close enough to the object $o$ to be able to observe its state. To do this the agent invokes  \Call{navigateTo}{$o$} to steer itself from its current position to the location of $o$. 
The travel to $o$ can be done by implementing a graph-based path-finding algorithm such as A* \cite{hart1968formal,millington2019AI}. 

\begin{algorithm}
\small
\caption{\em For selecting and executing tasks, parameterized by two heuristics.}\label{alg:solver}
\begin{algorithmic}[1]
\Procedure{Solver}{$selectH, findH$} \label{solver.alg.header}
\While{$budget > 0$}
\If{$toDo \not = \emptyset$}
   \State $T \gets selectH(toDo)$ \Comment{use the task-selection heuristic}
    \If{$T \neq null$} 
        \State $(o,\psi, S) \gets T$
       $\label{search.invoke.selectNode}$
       \State $toDo \gets toDo/\{ T \}$
       
        \State \Call{navigateTo}{$o$}  \label{search.alg.navigate}
        \If{$\neg \psi$} 
            \Call{dynamicGoal}{$o,\psi,S, findH$} 
            \label{search.alg.dyngoal1}  
        \EndIf
        \If{$\psi$} 
          \  $done \gets done \cup \{ T \}$
        \Else \ $toDo \gets toDo \cup \{ T \}$
        \EndIf
    \EndIf    
\Else
   \If{there is terrain unexplored} \Call{findTask}{ } 
    \label{search.alg.explore}   
   \Else \ \Return
   \EndIf
\EndIf
\EndWhile
\EndProcedure
\end{algorithmic}
\end{algorithm}

When the agent reaches $o$, and its state satisfies $\psi$ the testing task $T$ is completed with a positive result (pass). That is, it is confirmed that the state $\psi$ is reachable. If the state of $o$ at that moment does not satisfy $\psi$ the agent needs to do something to change the state of $o$.
A quite common logic in games is that toggling the state of an object $q$ requires 
another object $i$ to be interacted (so-called 'enabler' in Section \ref{problem setup}). We do not assume that the agent has pre-knowledge of which object is the enabler (or enablers, if there are more than one) of a particular target $o$. So, it invokes the algorithm \Call{dynamicGoal}{} to search for such an enabler.


\subsubsection*{\Call{dynamicGoal}{$o, \psi, S, findH$}} This procedure can be thought of as deploying a goal to change $o$ to a state satisfying $\psi$. 
The procedure first calculates a set $\Delta$ of objects that have been seen by the agent
and have not been 'tried' before (line \ref{alg:dynamicGoal:delta}). These are candidate objects, whose interaction might change the state of the target $o$.
An object $i \in \Delta$ is selected based on the  $findH$ heuristic. For example, the heuristic might favor objects of a certain type, or it might favor closer $i$. When $i$ is selected, it is marked to avoid choosing it again.
If $findH$ cannot come up with an $i$, e.g. because $\Delta$ is empty, the agent explores the game, searching for a more objects.  
If we have an $i$, the agent navigates to it. Once $i$ is reached, an appropriate action (such as picking up or interacting) is performed on $i$. In order to determine whether $o$ has changed and $\psi$ is met, the agent navigates back to $o$.
Some coordination is also needed during these steps. The object $i$ is critical towards checking $\psi$. That is, other agents should refrain from messing with $i$. To enforce this, we employ locking on $i$, which is released again after the checking of $\psi$.
If $\psi$ is established, \Call{dynamicGoal}{ } is done. Else, the process of calculating $\Delta$ and selecting (another) $i$ is repeated. 
This goes on either until $\psi$ is verified, or the stop condition $S$ becomes true (and then the verification verdict would be a 'fail'). 

\begin{figure}[h]
\centering
\includegraphics[width=\linewidth,height=5cm,scale=0.5]{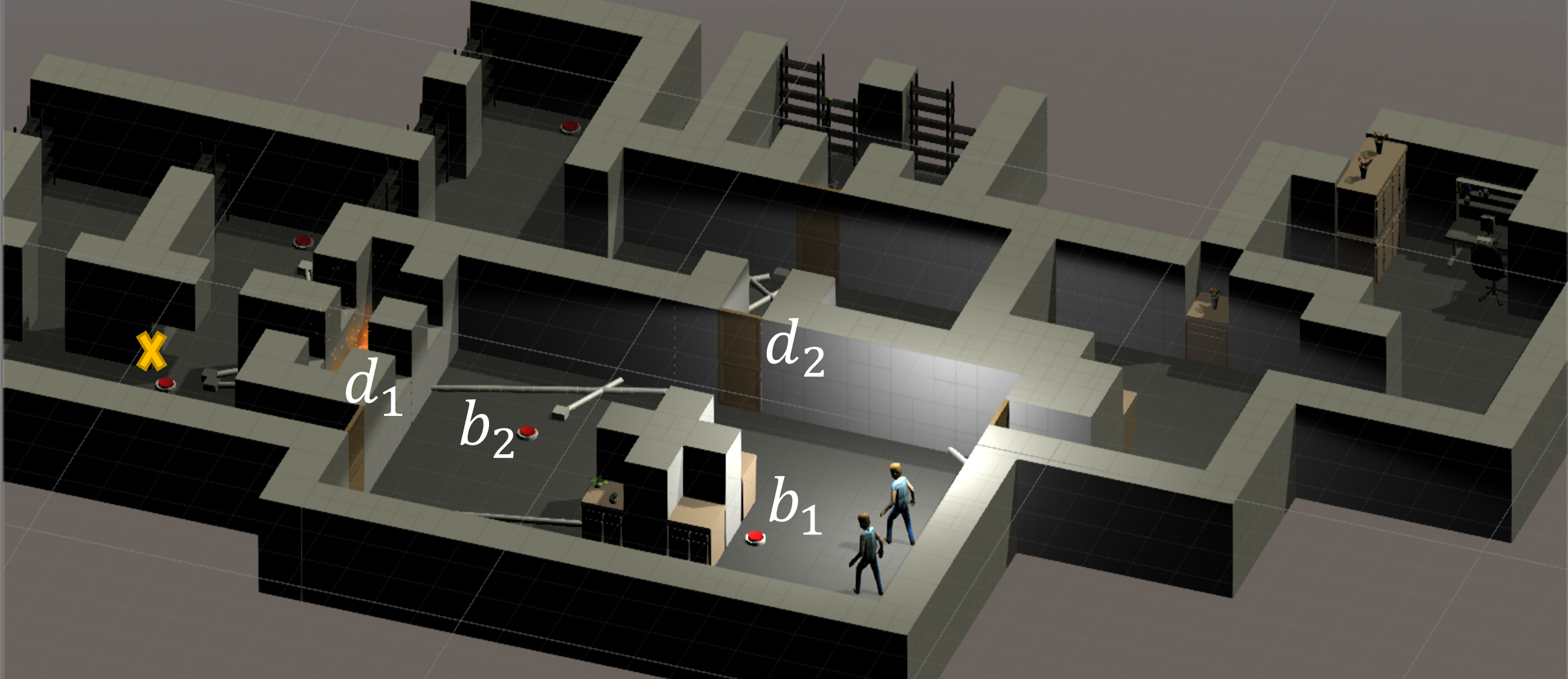}
\caption{\rm\em Screenshot of the game Lab Recruit.}
\label{fig.LRFinal}
\vspace{-3mm}
\end{figure}

To give a more precise example of when to deploy a $dynamicGoal$, consider the following testing task:

\begin{example} \label{example.T1} 
Consider the door $d_1$ in the game in Fig. \ref{fig.LRFinal}.
$T_1 = (d_1, d_1.open = true, S)$ where $S$ is some stop condition. 

\end{example}


Suppose an agent shown in Fig. \ref{fig.LRFinal} takes the task $T_1$. Since $d_1$ is closed at that moment, a dynamic goal to open $d_1$ is added (a call to \Call{DynamicGoal}{}). This would then try buttons $b_1$ and $b_2$ (see Fig \ref{fig.LRFinal}) to make $d_1$ open. It will not try a button twice, as it marks buttons it has touched.

\begin{algorithm}
\footnotesize
\caption{\em For solving a single task, parameterized by one heuristic.}\label{alg:dynamicGoal}
\begin{algorithmic}[1]
\Procedure{DynamicGoal}{$ o, \psi, S, findH$}
   \While{$\neg \psi \wedge \neg S$}
   \State $\Delta \gets \{ i \: | \; i \in seenObj \wedge  \neg mark_o(i) \wedge \neg locked(i) \}$
   \label{alg:dynamicGoal:delta} 
   \State choose $i \in \Delta$, based on $findH$
   \Comment{use the object-selection heuristic}
   \If{ $i = null$}
       \If{there is terrain unexplored}
            \State \Call{explore}{ }   \Comment{explore world to find new objects}
       \Else
            \State \Return
       \EndIf
    \Else 
    \State $mark_{o}(i) \gets true$ \Comment{mark $i$ as tried for $o$}
        \State \Call{lock}{$i$}
        \State $navigateTo(i)$ 
        \State $applyAction(i)$ \Comment{such as interact}
        \State $navigateTo(o)$ 
        \State \Call{unlock}{$i$}
   \EndIf
   \EndWhile
\EndProcedure
\end{algorithmic}
\end{algorithm}

Back in \Call{Solver}{}, the agent uses Algorithm \ref{alg:exploration}  \Call{findTask}{} when the $toDo$ set is empty (there is no open task, whose location of its target object is known). The agent explores the game world to learn the game's spatial layout, finding game objects as it goes. A graph-based exploration algorithm such as \cite{prasetya2020navigation} can be used. 
It stops upon finding some new target objects (objects targeted by tasks), or if there is no terrain left to explore.
The overall exploration heuristics may affect the overall task-solving performance. A few examples are mentioned below:

\begin{itemize}
    \item {\em Gradually}: the agent explores the world only until it sees new target objects. Algorithm \ref{alg:exploration} does this.
    
    \item {\em Aggressive}: 
    as Algorithm \ref{alg:exploration}, but the agent does not stop until there is nothing left to be explored. 
    
    \item {\em Limited budget}: as Algorithm \ref{alg:exploration}, but the algorithm is given a certain budget, and stops when the budget runs out.
\end{itemize}

\begin{algorithm}
\small
\caption{\em For finding new tasks.}\label{alg:exploration}
\begin{algorithmic}[1]
\Procedure{findTask}{ }
\While{there is terrain unexplored}        
    \State \Call{basicExplore}{ }
    \State $V \gets$ newly observed objects
    \State $W \gets \{ (o,\psi,S) \; | \; o \in V \wedge (o,\psi,S) \in \mathcal{T}/(toDo \cup done) \}$
    \If{$W \neq \emptyset$}
       \State $toDo \gets toDo \cup W$
       \State \Return 
    \EndIf
\EndWhile
\EndProcedure
\end{algorithmic}
\end{algorithm}

\subsubsection{Synchronization Level} 

To later investigate the impact of information sharing and synchronization between the agents, we consider two levels of sharing: 

\begin{itemize}
    \item {\em Basic}: the agents share seen tasks and solved tasks.
    Algorithm \ref{alg:multiAgents} already does this, by maintaining common $toDo$ and $done$ sets. Additionally, the agents also share the {\em location} of tasks' target objects. 
    
    \item {\em Extended}:
    this extends the basic sharing above by having the agents to also share information about explored areas to each other, e.g. information about discovered navigation mesh and object states. 
    As common in a multi-agent setup, each agent has its own state. So, sharing information involves sending the information from one agent to another, and synchronizing the sent information into the receiving agent's own belief. This incurs some computation overhead.
 
\end{itemize}


\section{Experiments}\label{sec:experiment}

This section discusses a series of experiments aimed to investigate the following research questions:

\begin{itemize}
    \item {\bf RQ1:} {\em does multi-agent  speed up testing?}
    \item {\bf RQ2:} {\em how well can multi-agent deal with complex logic?}
\end{itemize}

We implement the agents using iv4XR\footnote{\url{https://github.com/iv4xr-project/aplib}}, a Java multi-agent programming framework with a particular focus on game testing \cite{prasetya2020aplib}. The purpose of an iv4XR agent is to control an in-game entity; 
for instance to control a player character of a game. 
As such, the agent can interact with  game and control it  just like a human player can. The framework is inspired by the popular Belief-Desire-Intention concept of agency \cite{herzig2017bdi}, where an agent has {\em belief},  representing information the agent has about its current environment. 
Iv4XR provides automated world navigation and exploration algorithms \cite{prasetya2020navigation}, and test agents can be equipped with them. Having such a feature enables us to define testing tasks at a more functional level, allowing us to abstract away details related to, for example, physical 3D navigation.  In our multi-agent approach, each agent simulates a player and responds dynamically to the game under test. 

For the experiments, a multi-player 3D game called Lab Recruit (LR) is used\footnote{\url{https://github.com/iv4xr-project/labrecruits}}. Figure \ref{fig.LRFinal}  shows an example screenshot of the game. 
LR allows new game {\em levels}\footnote{In gaming, a 'level' refers to a world or a maze, playable within the same game. To extend the play-value of a game, developers often provide a set of levels.}  to be defined using plain text files. 
This allows us to generate a range of various levels for our experiments.
The playing goal of an LR level can simply be to explore it, or to reach a certain end-room. 
There are two types of game objects in LR that are of particular interest for the experiments: doors and buttons.
A door guards the access to the rooms it connects. 
The state of a door can be altered (e.g. from closed to open) by interacting with a button that is connected to it. The players do not know upfront which buttons are connected to which doors.
Furthermore, the players' have limited view range.
Throughout the game, players 
can be thought to gain points by going through rooms (e.g. due to certain items in the rooms). Some rooms give much more points, and are thus important for the game play.  
Verifying that all doors in a level can be opened and reachable would prove the level's basic correctness. 
If time is limited, verifying the doors guarding high valued rooms can be considered as more important.


Several factors affect the performance a multi-agent setup, such as 
the size of the game levels and the task distribution among agents.
To investigate them different LR levels are created. They are all variations of a level we will call $Basic{\text -}Level$. 
%
%
This is a level with a
$10m{\times} 10m$ main hall
with ten doors (blockers) guarding access to side rooms. There is a button in front of each door that opens the door.
Six of the doors have in-game points of one, and four have in-game points of ten (they are guarding high-value rooms).
The higher point doors are $d_2$, $d_3$, $d_6$, and $d_9$.

\textbf{Testing tasks}. In the experiments, we consider a set of testing tasks, one $T_d$ for each door $d$ in the target level, to verify that a state where the door $d$ is open is reachable.
$Basic{\text -}Level$ has thus ten testing tasks.

\textbf{Heuristics setting}.
Recall that each agent runs 
the algorithm \Call{Solver}{}. It is parameterized by two heuristics: $selectH$ and $findH$. 
For the task selection heuristic ($selectH$), three heuristics are taken into consideration: one that chooses tasks randomly, one that targets tasks whose value is higher than a given threshold, and one that targets tasks with values below a threshold.
Agents that use them are referred to as $A_R$, $A_H$ and $A_L$ respectively.
The $findH$ heuristic (to choose a candidate enabler to try) is set to choose a button that is closest to the agent's position. 
The heuristic for \Call{findTask}{} is set to be the same as what is already in its algorithm, namely to explore gradually.

\textbf{Example.} Take a look at the level depicted in Fig. \ref{fig.LRFinal}. Suppose the agents initially observe nothing because of their limited visibility range. They then begin to gradually explore the level. 
Imagine that they see $d_1$ and $d_2$. Since these are targets of testing tasks, the corresponding tasks $T_{d1}$ and $T_{d2}$ are added to the $toDo$ set. Both are worth one point. Since the testing task is not empty, the agents first look through the $toDo$ to see if any tasks are available before choosing one. 
Suppose one of the agents is $A_L$. 
The two tasks in $toDo$ are of low value, so they are targets for $A_L$. Imagine that $T_{d1}$ is chosen. 
The agent $A_L$ then navigates to $d_1$. As it is closed, $A_L$ invokes \Call{DynamicGoal}{} to open it. 
Suppose at that point $A_L$ has seen the button $b_1$, and it also happens to be located nearest to $A_L$.
By the heuristic $findH$, the agent then chooses $b_1$, to be tried (interacted) in order flip the state of $d_1$ to open. If this happens and confirmed (by travelling to $d_1$ to confirm its state), task $T_{d1}$ is completed and 
the agent can proceed to the next testing task. 

All experiments were run on a Windows machine equipped with an Intel(R) i7-8565U (4 cores) processor at 2.8 GHz  and 32GB RAM. Every run in the experiment is repeated three times, and the result is the average of the runs. 

\subsection{RQ1: does multi-agent speed up testing?}

To answer the RQ, we consider a number of factors that affect performance:

\begin{itemize}
    

    \item Information synchronization. The two synchronization levels, basic and extended, mentioned in Section \ref{sec:multi-agent-algorithm} will be considered.

    \item Different team compositions, consisting of agents with different task selection heuristics ($selectH$), will be considered. 

    \item View distance: agents' limited visibility complicates exploration and task finding. We would like to investigate how the performance is affected if a larger view distance is allowed. 
    



\end{itemize}

\subsubsection{Effect of information synchronization}

We run our multi-agent setup on ten levels, variations of $Basic{\text -}Level$ with increasing size. For each level, we deployed a team of two agents, namely $\{ A_H, A_L\}$. Two information synchronization levels are tried: basic and advanced. 
The agents' visibility range is set to six. 

\begin{figure}[h]
\centering
\includegraphics[width=\linewidth,height=5cm,scale=0.5]
{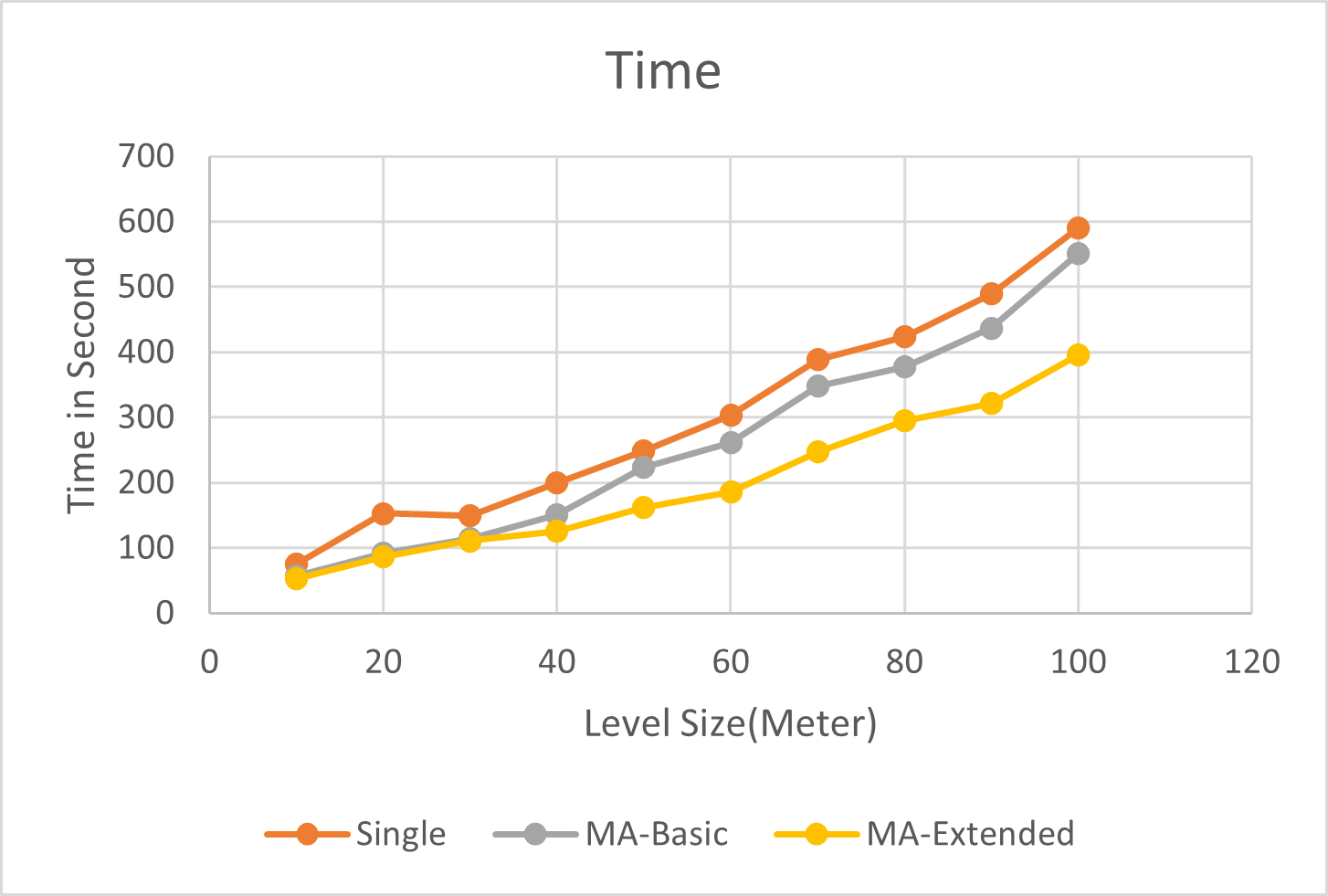}
\caption{\rm\em The time needed to solve all the tasks. 
The $x$-axis is the level size, e.g. $20{\times}20$.
The $y$-axis is time in second. 
$\rm Single$ uses just one agent.
$\rm MA{\text -}Basic$ and $\rm MA{\text -}Extended$ use two agents, using the basic and respectively extended information synchronization.}
\label{fig.multi-single-time}
\vspace{-3mm}
\end{figure}

Fig. \ref{fig.multi-single-time} shows the results. 
The multi-agent setup with extended synchronization outperforms the single-agent setup, in particular in bigger levels, despite the overhead of having to synchronize information.
E.g. it is nearly 35\% faster than a single agent on the $100{\times} 100$ level. 
When an agent chooses a task whose target object $o$ it has never seen before, the shared navigation information from the other agent may help it to find $o$, saving time that is otherwise needed for exploring the world in order to find it. In contrast, basic synchronization does share the locations of target objects, but not the path to them. So in this case the agent still has to search the level to actually get to $o$. 
Using only basic synchronization, the multi agent setup is still faster than a single agent, but the results clearly show that using extended synchronization, despite its overhead, pays off.

\begin{figure}[h]
\centering
\includegraphics[scale=0.6]{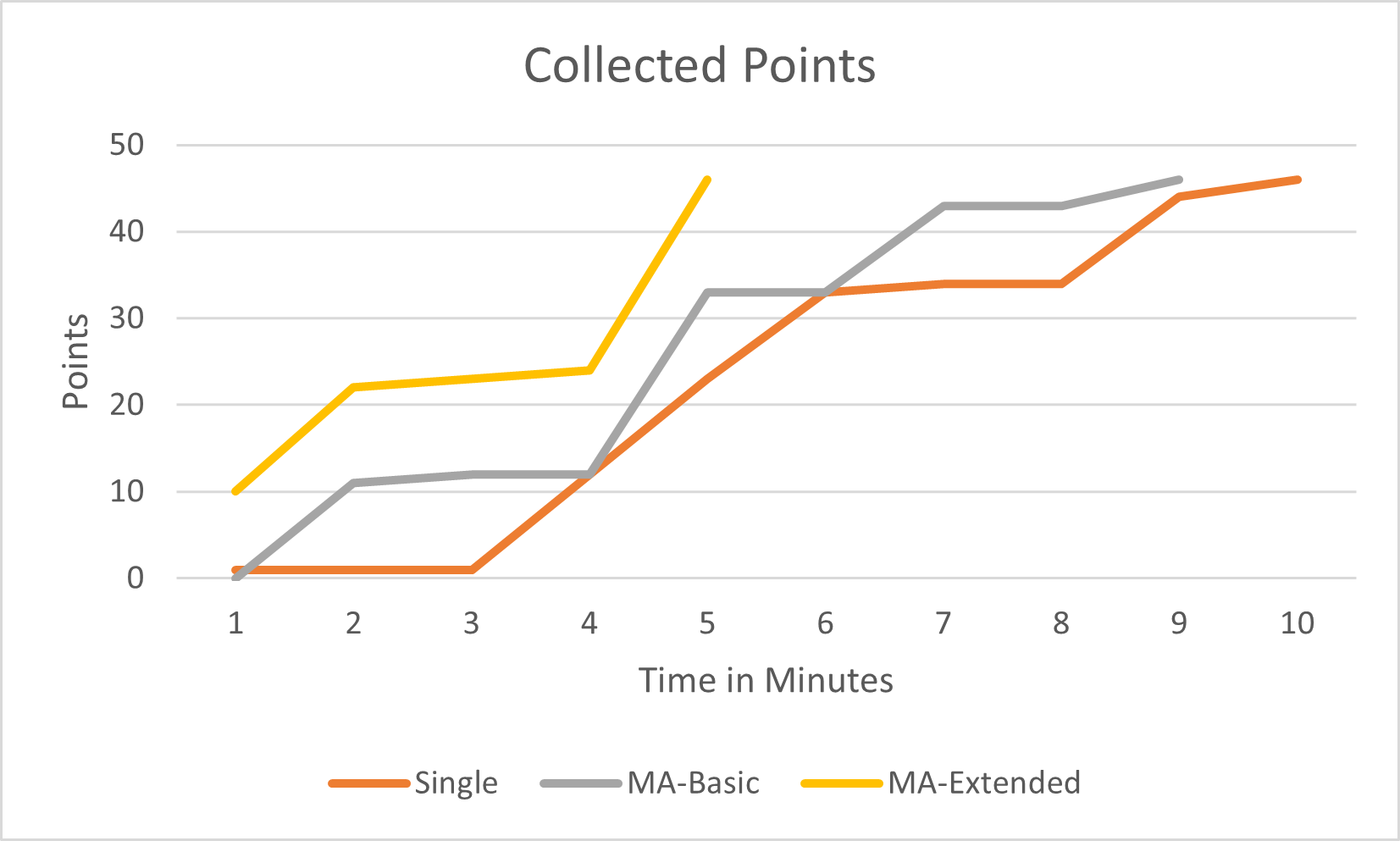}
\caption{\rm\em Points collected over time on a level with size $100{\times}100$. Collected points correlate with the number of done tasks.}
\label{fig.sharedPoint}
\vspace{-3mm}
\end{figure}

Fig. \ref{fig.sharedPoint} shows the points collected after a specific time on the $100{\times}100$ level. 
Recall that opening a door gives points, so the collected total points correlate with the number of tasks completed. A steeper increases in the graph corresponds to the completion of the verification of a high-valued door, which is more important.
When a total of 46 points is reached, all tasks are completed. 
Multi-agent with extended synchronization can complete all the tasks in five minutes, while a single agent can only complete two important tasks.

\subsubsection{Different team composition}

In this experiment, we run different teams of agents on a $100{\times}100$ level.
A team consists of agents with different task selection heuristics. 
In addition to the aforementioned $A_H$, $A_L$, and $A_R$ agents we add $A_{EX}$ and $A_{E}$.
Agent $A_{EX}$ does not select any tast. It just explores the world to discover tasks' target objects to help other agents by sharing information.
Agent $A_E$ is an eager agent that takes a task as soon as it sees its target object.

\begin{figure}[h]
\centering
\includegraphics[scale=0.6]{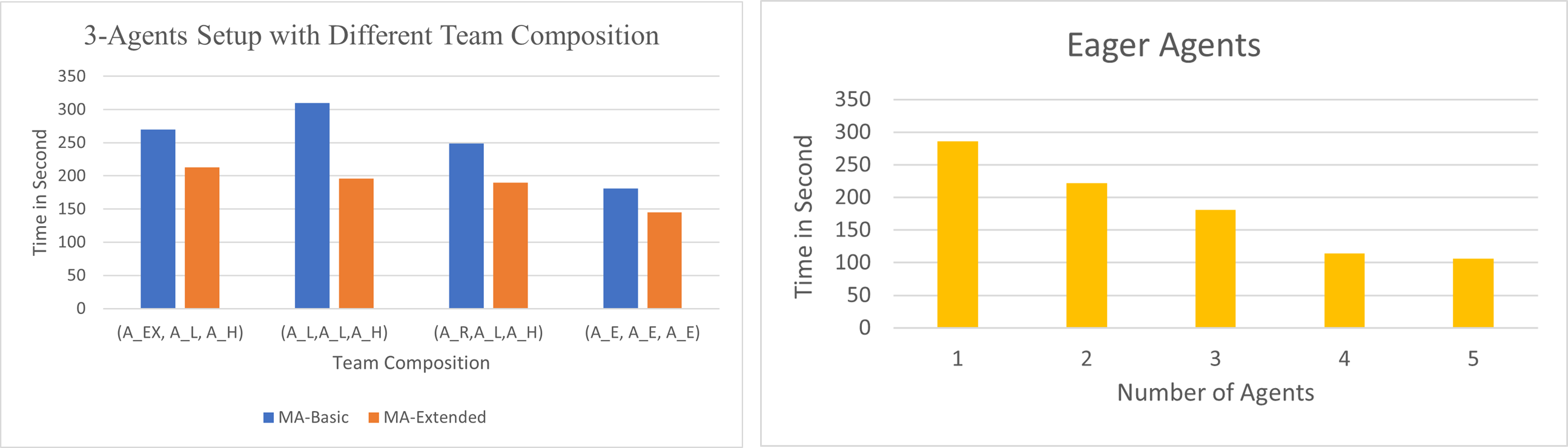}
\caption{\rm\em Three agents setup with different task selection heuristics on a $100{\times}100$ level. In addition to $A_R$, $A_L$ and $A_H$, we have 
$A_{EX}$ and $A_E$; these are agents with exploration and eager heuristics, respectively.
In the graph on the right on the right side, only eager agents are used, but with one up to five of them. }
\label{fig.3Agents}
\vspace{-3mm}
\end{figure}

Fig. \ref{fig.3Agents}-left shows the results for different teams of three agents. 
%
%
The results show that including a dedicated exploration agent in a team (team $(A_{EX}, A_L, A_H)$) does not really improve the performance. This is because the agents still need to do the tasks. On the other hand, the performance is {\em not} significantly worse either, despite having one less worker agent, which shows that the shared information from $A_{EX}$ does help.

We also conducted an experiment to examine the impact of team size. 
Fig. \ref{fig.3Agents}-right shows the performance of teams consisting of one up to five eager agents.
Only eager agents are used because in the previous experiment they have the best performance.
The outcome indicates that performance gets better as the number of agents increases. When there are five agents instead of only one, the testing time is reduced by nearly one-third and by more than half when there are two agents. 
The performance of five agents is not much different from four agents, however. A possible reason is because we have only have four CPU cores in our experiment setup.

\subsubsection{View distance}

\begin{figure}[h]
\centering
\includegraphics[scale=0.65]{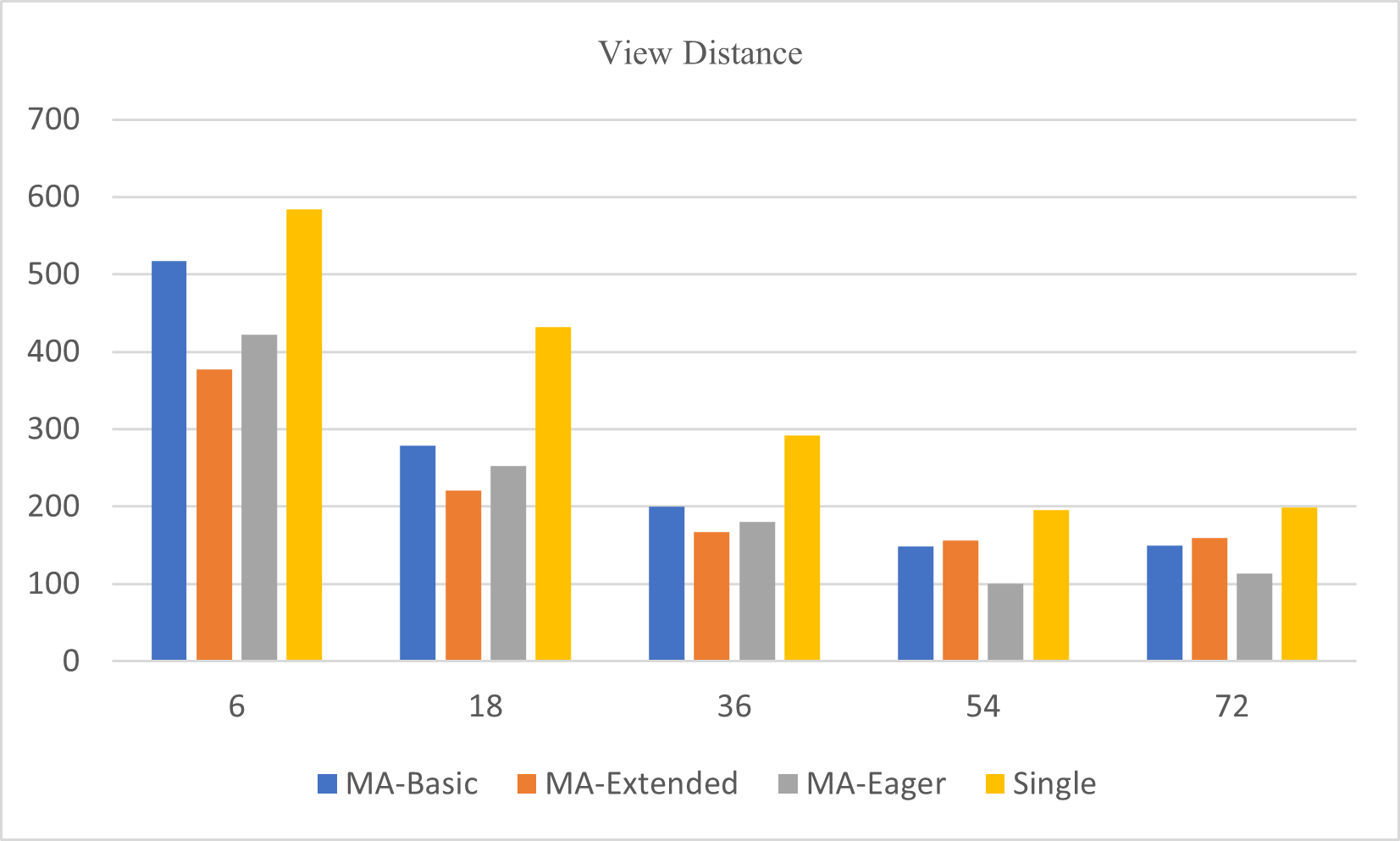}
\caption{\em 
The performance of different setups under 
different view distance on a $100{\times}100$ level. {MA-Basic} and {MA-Extended} use two agents $A_H$ and $A_L$ with basic and extended synchronization, respectively.
{MA-Eager} uses two eager agents with advanced synchronization.
}
\label{fig.viewDistanceWith3Setup}
\vspace{-3mm}
\end{figure}

To investigate the impact of view distance we run a setup with two agents on a $100{\times}100$ level, with varying view distance.
Fig. \ref{fig.viewDistanceWith3Setup} shows the results.
It shows that when the view distance increases, test performance improves. 
This is as expected, as by increasing the view distance, the agents can see more objects/tasks and can choose a task based on how far it is from its position. 
Multi-agent setups consistently outperforms the single-agent setup, which is also expected.
However, when the view distance increases, the performance of multi-agent setups using various strategies is nearly equal. 
Also, at some point further increase of the view distance does not significantly alter performance. The reason is that, even though the locations are known, the agent still needs to navigate to each task to solve it, and this costs time.

It should be noted that enlarging the view distance may not be an option provided by the game under test, e.g. to keep the amount of data that the game needs to send over to the agents light weight. 

\subsection{RQ2: how well can multi-agent deal with complex logic?}


Game logic might make solving a testing task more challenging as the agents do not have full pre-knowledge about the logic. 
For RQ2 we will consider a number of door-logics, listed below, which are increasingly more challenging. These logics are quite common in for example RPG games.

\begin{itemize}
    \item {\em Distant-connection logic}.
    A door has this logic if its enabler is located far from it. 
    No door in $Basic{\text -}Level$ has far enabler. So in the experiments later, we will create variations of the level.
    
    \item {\em Chained-connection logic.}
    A door $d$ has this logic if its enabler is 'hidden' in another room, guarded by its own door.
    So, to open $d$ multiple enablers have to be interacted, and in a specific order.
    It gets more complicated if the chaining is deeper than one (deeper chains are not included in our experiments).
    
    \end{itemize}

\vspace{-5mm}
\subsubsection{Distant-connection logic}
To investigate how well a multi-agent setup can deal with the distant-connection logic, a variation of a $30{\times}30$ $basic{\text -}level$ is used.
Some of the doors are changed to have a distant connection logic, by placing its corresponding button in a randomly far location.
We created five different instances of $basic{\text -}level$ with different numbers of such doors, 
starting from two to ten. We compare a single agent with a team of two agents $\{ T_H, T_L\}$, using either basic or extended synchronization. 

Table \ref{tab.randomconnection} shows the result. 
It shows that when the number of distant enablers increases, the time needed to finish the task increases, too. Even with extended synchronization, it becomes harder for agents to solve all the tasks.

Multi-agents with advanced synchronization can perform better than the other two other setups, except when all doors in the level have the distance connection logic,
where the single agent setup is unexpectedly superior.
The likely cause is accidental tasks solving. This happens when an agent unintentionally completes a task that is not intended for it. This occurs when the agent interacts with $i'$ that is linked to $d'$ while searching for an enabler to open a door $d$.
The task $T_{d'}$ is then declared as completed, though $T_d$ remains uncompleted. If the door is open when the agent sees it, it will be considered completed and removed from the testing tasks.


\begin{table}
  \caption{\rm\em The performance on a setup with doors with distant connection logic.} \label{tab.randomconnection}
  {\small
  \begin{center}
  \vspace{-4mm}
  \begin{tabular}{ccccl}
    \toprule
    \#DC & Single & MA-Basic &  MA-Extended\\
    \midrule
    2 &  210  & 159 & 117  \\
    4 & 265 & 209 &   195  \\    
    6 & 315 & 289 &   215 \\
    8 & 352 &   333 & 232  \\
    10 & 180 &   352 & 264  \\
    \bottomrule
  \end{tabular}
  \end{center}
  }
  \vspace{-6mm}
\end{table}

\vspace{-5mm}
\subsubsection{Chained-connection logic} 
Fig. \ref{fig.LRFinal} shows an example of a level with a 'hidden' button, marked by X. This means that to open $d_2$, $d_1$ must first be opened. If the agents take the tasks in such a way that $d_2$ is to be opened by agent $A$ and $d_2$ by agent $B$, this presents a challenge. Until $B$ opens $d_1$, $d_2$ cannot be opened. So, $A$ has to wait, and we lose parallelism. Note that the doors' logic is not known to the agents upfront, so it also not possible for the agents to know upfront what an ideal task distribution is. Three distinct levels are created, each with a unique arrangement of one, two, and three hidden buttons. Our comparison of the time required to complete all tasks is shown in Table. \ref{tab.hiddenButton}. 

The multi-agent setup with extended information synchronization outperforms the single-agent setup. Increasing the number of hidden buttons increases the overall testing time. Having three chained-connection doubles the testing performance.
As previously mentioned, one of the causes is the choices the agents make when selecting the tasks to do. The choices taken may force an agent to wait for another agent to access to certain areas, and hence the overall task solving becomes longer. Secondly, by attempting random buttons that are actually not connected to the door that guards a target room, the agent may block access to another area.  


\begin{table}
  \caption{\rm\em Different numbers of chained connections in a basic level with a size of 30.  
 } \label{tab.hiddenButton}
  {\small
  \begin{center}
  \vspace{-4mm}
  \begin{tabular}{ccccl}
    \toprule
    \#HB & Single & MA-Eager &  MA-Extended\\
    \midrule
    0 &  149 & 128 & 111   \\
    1 & 214 &   145 & 140   \\    
    2 & 307 &   254 & 191  \\
    3 & 337 & 308 &   241   \\
    \bottomrule
  \end{tabular}
  \end{center}
  }
  \vspace{-6mm}
\end{table}

\vspace{-8mm}
\section{Related Work}\label{sec:related-work}


Automated game testing is a challenging problem. The interaction space of a game is often very large and it is challenging for an algorithm to steer a game under test to get to a particular state that needs to be tested. 
Much of recent work on automated game testing has been focused on the use of AI \cite{zarembo2019analysis}. For example, the use of Monte Carlo Search Tree (MTCS) for generating play tests was studied in \cite{ariyurek2020enhancing,ariyurek2019automated}. The use of reinforcement learning was studied in e.g. \cite{pfau2017automated,zheng2019wuji,ferdous2022towards}.
These approaches require little human steering, though the training time could be excessive. 
Model-based games testing was studied in \cite{iftikhar2015automated,ferdous2021search}. These approaches use a behavior model e.g. in the form of an Extended Finite State Machine (EFSM), from which test cases are generated. Test generation is fast, but on the other hand an EFSM model is needed. The model needs to be refined enough to make sure that the generated test cases are actually executable. Crafting such a model is costly.

Most of the works mentioned above \cite{ariyurek2020enhancing,ariyurek2019automated,pfau2017automated,zheng2019wuji,ferdous2021search,ferdous2022towards} are arguably agent-based in a broad sense that they use a test agent to control the player character of the game under test. 
In our previous work, we studied the use of BDI agents
\cite{shirzadehhajimahmood2021using,prasetya2022agent}. E.g. these agents memorize the states of seen game objects, believing that they remain unchanged until proven otherwise. The agents also use path planning to navigate the world, based on  navigation information they have in their belief. This eliminates the need for expensive training as in e.g. reinforcement leaning, though on the other hand, additional implementation effort may be needed to enable a navigation graph to be constructed automatically during the tests.

The work in \cite{ferdous2021search} combines model-based testing, search-based testing (SBT), and agent-based testing. An SBT algorithm is used to generate test-cases from an EFSM model. The model in \cite{ferdous2021search} can remain quite abstract. The approach exploits a BDI agent as a smart executor of test sequences generated from such an abstract model.
The work in \cite{ferdous2022towards} combines reinforcement learning and BDI agents. The BDI layer is used to provide an abstract concept of actions (e.g. to auto-navigate to a given target object), so that the reinforcement learning part only needs to deal with such abstract actions.

Most of the approaches mentioned above are single agent. 
The works in \cite{gordillo2021improving,zheng2019wuji} use multiple agents, though these are individual agents that run in parallel to train a {\em common} model, or a population of common models. 
Arguably this is a form of multi-agent cooperation, but not in the sense that the agents directly cooperate with each other. 
The latter, so the use of cooperating agents for game testing, has not been much studied.




\section{Conclusion and Future Work} \label{sec:concl}

\vspace{-3mm}
We presented a cooperative multi-agent testing approach, targeting mainly puzzle-based and world-exploration games and evaluated the approach using a case study of a 3D game called Lab Recruits.
Basic and extended  information synchronisation were considered, as different levels of cooperation. In the latter, agents share information about explored areas to each other in addition to sharing the location of testing tasks' target objects which is shared at the basic level. 
We evaluated the differences in test performance between a single-agent and multi-agent setups. The experiment demonstrates that multi-agent with extended information sharing consistently performs better than single agent and multi-agent with just basic sharing. Adding more agents was demonstrated to improve performance. We also demonstrate that multi-agent can handle complex game logic, while still being superior to a single agent.
The study was done with one case study. For future work we would like to investigate more case studies, e.g. more games or considering more variations in  the layout of the game world.
Also, currently information synchronization is done by greedily pushing the information to share. We can consider a more lazy pull mechanism. It is more complex, but may give further performance improvement. This is future work.

%
%
%
\bibliographystyle{splncs04}
\bibliography{Bibliographies}

\begin{thebibliography}{10}
\providecommand{\url}[1]{\texttt{#1}}
\providecommand{\urlprefix}{URL }
\providecommand{\doi}[1]{https://doi.org/#1}

\bibitem{albaghajati2020video}
Albaghajati, A.M., Ahmed, M.A.K.: Video game automated testing approaches: An assessment framework. IEEE transactions on games  (2020)

\bibitem{ariyurek2019automated}
Ariyurek, S., Betin-Can, A., Surer, E.: Automated video game testing using synthetic and human-like agents. IEEE Transactions on Games  (2019)

\bibitem{ariyurek2020enhancing}
Ariyurek, S., Betin-Can, A., Surer, E.: Enhancing the monte carlo tree search algorithm for video game testing. In: 2020 IEEE Conference on Games (CoG). IEEE (2020)

\bibitem{ch2007agent}
Ch.~Meyer, J.J.: Agent technology. Wiley Encyclopedia of Computer Science and Engineering pp.~1--8 (2007)

\bibitem{ferdous2021search}
Ferdous, R., Kifetew, F., Prandi, D., Prasetya, I., Shirzadehhajimahmood, S., Susi, A.: Search-based automated play testing of computer games: A model-based approach. In: International Symposium on Search Based Software Engineering. pp. 56--71. Springer (2021)

\bibitem{ferdous2022towards}
Ferdous, R., Kifetew, F., Prandi, D., Susi, A.: Towards agent-based testing of {3D} games using reinforcement learning. In: 37th {IEEE/ACM} International Conference on Automated Software Engineering (2022)

\bibitem{gordillo2021improving}
Gordillo, C., Bergdahl, J., Tollmar, K., Gissl{\'e}n, L.: Improving playtesting coverage via curiosity driven reinforcement learning agents. In: 2021 IEEE Conference on Games (CoG). pp.~1--8. IEEE (2021)

\bibitem{hart1968formal}
Hart, P.E., Nilsson, N.J., Raphael, B.: A formal basis for the heuristic determination of minimum cost paths. IEEE transactions on Systems Science and Cybernetics  \textbf{4}(2),  100--107 (1968)

\bibitem{herzig2017bdi}
Herzig, A., Lorini, E., Perrussel, L., Xiao, Z.: Bdi logics for bdi architectures: old problems, new perspectives. KI-K{\"u}nstliche Intelligenz  \textbf{31}(1),  73--83 (2017)

\bibitem{iftikhar2015automated}
Iftikhar, S., Iqbal, M.Z., Khan, M.U., Mahmood, W.: An automated model based testing approach for platform games. In: 2015 ACM/IEEE 18th International Conference on Model Driven Engineering Languages and Systems (MODELS). pp. 426--435. IEEE (2015)

\bibitem{liu2022prospects}
Liu, Y., Li, Z., Jiang, Z., He, Y.: Prospects for multi-agent collaboration and gaming: challenge, technology, and application. Frontiers of Information Technology \& Electronic Engineering  \textbf{23}(7),  1002--1009 (2022)

\bibitem{millington2019AI}
Millington, I., Funge, J.: Artificial intelligence for games, 3rd edition. CRC Press (2019)

\bibitem{ostrowski2013automated}
Ostrowski, M., Aroudj, S.: Automated regression testing within video game development. GSTF Journal on Computing  \textbf{3}(2) (2013)

\bibitem{pfau2017automated}
Pfau, J., Smeddinck, J.D., Malaka, R.: Automated game testing with icarus: Intelligent completion of adventure riddles via unsupervised solving. In: Extended Abstracts Publication of the Annual Symposium on Computer-Human Interaction in Play. pp. 153--164 (2017)

\bibitem{prasetya2020aplib}
Prasetya, I.S.W.B., Dastani, M., Prada, R., Vos, T.E., Dignum, F., Kifetew, F.: Aplib: Tactical agents for testing computer games. In: International Workshop on Engineering Multi-Agent Systems. pp. 21--41. Springer (2020)

\bibitem{prasetya2022agent}
Prasetya, I., Pastor~Ric{\'o}s, F., Kifetew, F.M., Prandi, D., Shirzadehhajimahmood, S., Vos, T.E., Paska, P., Hovorka, K., Ferdous, R., Susi, A., et~al.: An agent-based approach to automated game testing: an experience report. In: 13th International Workshop on Automating Test Case Design, Selection and Evaluation (2022)

\bibitem{prasetya2020navigation}
Prasetya, I., Voshol, M., Tanis, T., Smits, A., Smit, B., Mourik, J.v., Klunder, M., Hoogmoed, F., Hinlopen, S., Casteren, A.v., et~al.: Navigation and exploration in 3d-game automated play testing. In: Proceedings of the 11th ACM SIGSOFT International Workshop on Automating TEST Case Design, Selection, and Evaluation. pp.~3--9 (2020)

\bibitem{schatten2017automated}
Schatten, M., Duri{\'c}, B.O., Tomi{\v{c}}i{\v{c}}, I., Ivkovi{\v{c}}, N.: Automated {MMORPG} testing --an agent-based approach. In: International conference on practical applications of agents and multi-agent systems. Springer (2017)

\bibitem{shirzadehhajimahmood2022online}
Shirzadehhajimahmood, S., Prasetya, I., Dignum, F., Dastani, M.: An online agent-based search approach in automated computer game testing with model construction. In: 13th International Workshop on Automating Test Case Design, Selection and Evaluation (2022)

\bibitem{shirzadehhajimahmood2021using}
Shirzadehhajimahmood, S., Prasetya, I., Dignum, F., Dastani, M., Keller, G.: Using an agent-based approach for robust automated testing of computer games. In: Proceedings of the 12th International Workshop on Automating TEST Case Design, Selection, and Evaluation. pp.~1--8 (2021)

\bibitem{zarembo2019analysis}
Zarembo, I.: Analysis of artificial intelligence applications for automated testing of video games. In: Proceedings of the 12th International Scientific and Practical Conference. Volume II. vol.~170, p.~174 (2019)

\bibitem{zheng2019wuji}
Zheng, Y., Xie, X., Su, T., Ma, L., Hao, J., Meng, Z., Liu, Y., Shen, R., Chen, Y., Fan, C.: Wuji: Automatic online combat game testing using evolutionary deep reinforcement learning. In: 2019 34th IEEE/ACM International Conference on Automated Software Engineering (ASE). pp. 772--784. IEEE (2019)

\end{thebibliography}

\end{document}